\documentclass[sigconf,nonacm]{acmart}

\usepackage{xspace}
\usepackage{soul}
\usepackage{paralist}
\usepackage{listings}
\usepackage{courier}
\usepackage{enumitem}
\usepackage{hyphenat}
\usepackage{bm}
\usepackage{cleveref}

\usepackage[T1]{fontenc}
\newcommand\shortdots{\makebox[1em][c]{.\hfil.\hfil.}}

\newcommand\etal{\emph{et al.}\xspace}

\def\BibTeX{{\rm B\kern-.05em{\sc i\kern-.025em b}\kern-.08emT\kern-.1667em\lower.7ex\hbox{E}\kern-.125emX}}

\acmConference[]{}{}{}

\begin{document}

\title{Are My Invariants Valid? A Learning Approach}

\author{Vincent J. Hellendoorn}
\email{vhellendoorn@ucdavis.edu}
\affiliation{%
  \institution{UC Davis}
  \streetaddress{1 Shields Ave}
  \city{Davis}
  \state{California}
  \postcode{95616}
}

\author{Premkumar T. Devanbu}
\email{ptdevanbu@ucdavis.edu}
\affiliation{%
  \institution{UC Davis}
  \streetaddress{1 Shields Ave}
  \city{Davis}
  \state{California}
  \postcode{95616}
}

\author{Oleksandr Polozov}
\email{polozov@microsoft.com}
\affiliation{%
  \institution{Microsoft Research}
  \streetaddress{14820 NE 36th Street}
  \city{Redmond}
  \state{Washington}
  \postcode{98052}
}

\author{Mark Marron}
\email{marron@microsoft.com}
\affiliation{%
  \institution{Microsoft Research}
  \streetaddress{14820 NE 36th Street}
  \city{Redmond}
  \state{Washington}
  \postcode{98052}
}
\renewcommand{\shortauthors}{Hellendoorn et al.}

\begin{abstract}
Ensuring that a program operates correctly is a difficult task in large, complex systems. Enshrining invariants -- desired properties of correct execution -- in code or comments can support maintainability and help sustain correctness.
Tools that can automatically infer and recommend invariants can thus be very beneficial. However, current invariant-suggesting tools, such as Daikon, suffer from high rates of false positives, in part because they only leverage traced program values from available test cases, rather than directly exploiting knowledge of the source code \emph{per se}.
We propose a machine-learning approach to judging the validity of invariants, specifically of method pre- and post-conditions, based directly on a method's source code. We introduce a new, scalable approach to creating labeled invariants: using programs with large test-suites, we generate Daikon invariants using traces from subsets of these test-suites, and then label these as valid/invalid by cross-validating them with held-out tests. This process induces a large set of labels that provide a form of noisy supervision, which is then used to train a deep neural model, based on gated graph neural networks. Our model learns to map the lexical, syntactic, and semantic structure of a given method's body into a probability that a candidate pre- or post-condition on that method's body is correct and is able to accurately label invariants based on the noisy signal, even in cross-project settings. Most importantly, it performs well on a hand-curated dataset of invariants.
\end{abstract}

\maketitle

\section{Introduction}
As software projects grow, their behavior gets harder to understand. Unexpected run-time values,
such as an index falling outside array bounds, can lead to crashes and breaches of security. For
this reason, developers often formalize invariants: statements about the domains of values, and the
relationships between them, that are expected during execution of the program. Whether about loops,
functions or classes, good invariants have many uses, including improving maintainability,
complementing test suites, simplifying debugging, and, if inserted into the code, detecting run-time
violations to ensure that faulty code ``fails fast."

Because invariants can be powerful safeguards, tools that can recommend useful invariants can be a boon to programmers. Daikon was introduced as a landmark tool that can suggest invariants based on run-time traces~\cite{ernst2007daikon}. It works by running a program's test suite (or realistic workloads) and tracking the values in the program to conjecture invariants about its variables and their relationships. This can include very specific statements, such as ``\verb+x+ is only stored once in this array"; if an invariant has enough supporting observations, it is presented to a programmer. Unfortunately, these invariants often do not hold in practice because of the incompleteness of trace data; we find that many of Daikon's invariants are false positives, especially its more complex statements.

A solution may lie in learning-based methods: a growing body of work has shown that program properties can be inferred with surprising accuracy from source code by supervised learners. For instance, machine learners and neural networks are able to accurately infer types for untyped code (such as JavaScript) based on just the code syntax~\cite{raychev2015predicting,Hellendoorn-DLTI,nl2type}. That this is possible can be explained by the high degree of ``naturalness'' in real-world code \cite{hindle2012naturalness}: due to the complexity of software development, developers tend to rely on highly repetitive, conventional patterns; this in turn allows the intended meaning to be easily inferred from the code itself. Examples of this phenomenon are the use of common identifiers and idioms \cite{Allamanis:2014:LNC:2635868.2635883,allamanis2014mining}; these patterns can be readily mined and exploited with natural language inspired models for a wide range of tasks \cite{ray2016naturalness, Hellendoorn-DLTI, allamanis2015suggesting,Allamanis:2014:LNC:2635868.2635883, Pradel:2018:DLA:3288538.3276517}, all using learned models that extract rich information from the code and/or its documentation. 

We conjecture in this paper that this can extend to program behavior as well: expectations about the behavior of methods can be inferred from the syntax of the method itself. However, while training data for variable names and types can be obtained at scale, correct invariants are difficult to obtain; they are fairly uncommon in code and using automated reasoning methods~\cite{sharma2013data} for labeling is not a very scalable proposition. At the same time, the task of learning to label invariants given source code is a complex one, requiring a powerful learner; therefore, to avoid over-fitting, we need lots of data. Thus, a learning approach is caught in a bind.

We introduce a novel approach out of this quandary, by creating a large (but noisy) labeled dataset: we feed
traces from subsets of test-suites into Daikon to produce candidate invariants and then label these
using the full test sets. While this approach is admittedly unsound, it is a way to obtain labels at
scale and could provide a sufficient training signal to train a powerful model, which can then be
tested on real, manually-curated data.

Our contributions are:
\begin{enumerate}
    \item We conduct a detailed manual evaluation of the validity of Daikon's invariants on a diverse set of projects, showing that most of its invariants are false positives.
    \item We propose a method for extracting both (likely) valid and invalid (but realistic) invariants at scale using a test suite, which can serve as a noisy, but useful training signal.
    \item We introduce a graph-based neural validator that is able to accurately discern which invariants are valid and analyze the characteristics of its performance.
    \item We show that our model achieves both high ROC scores on our mined invariant corpus and is able to accurately rank Daikon's invariants on our manually annotated dataset, contrary to several compared models.
\end{enumerate}

\section{Motivation}
\label{sec:motivation}
Much of a developer's time is spent reading code and reasoning about its run-time behavior.
Invariants aim to simplify this task by describing expectations about the values used in a program, both in isolation
(e.g. \lstinline{index >= 0}, or \lstinline{other != null}) and relative to each other (e.g.
\lstinline{end >= start}, or \lstinline{end == arr.Count}).
The most commonly used invariants, method pre- and post\hyp{}conditions,
specify the program's state before and after a method's invocation respectively.
Pre-conditions capture the expected state of the method's parameters and any relevant program state, whereas
post-conditions reflect how this state may have changed, as well as properties of the method's result.

Writing invariants, though, is tedious and often omitted.
As such, tools like Daikon offer a promising solution: automatically inferring method pre- and post-conditions just by
executing a test suite and monitoring the values in a program.
This approach is, however, inadequate in practice: Kim \& Petersen analyzed Daikon's behavior on several
programs and found its generated invariants mostly trivial or incorrect~\cite{kimevaluation}.
In other words, Daikon suffers from a high degree of false positives.

To better understand why, we decided to manually inspect invariants on real programs; we sampled 35 methods across seven large projects on Github and used Daikon to infer nearly 500 pre- and post-conditions for them.
We then had two raters hand-annotate these invariants (our precise methodology is detailed in \Cref{sec:golden}).
Among others, we found that only 30\% of the presented invariants were valid; 40\% were incorrect entirely and another
30\% were in no way relevant to the method.\footnote{This is after removing invariants on system constants, which
spanned over 20\% of initial invariants, see \Cref{sec:golden}.} However, the real story lies beyond the numbers: our
analysis painted a complex picture of invariants in modern software engineering.

\begin{lstlisting}[language={[Sharp]C}, caption={Method example from Metrics.Net \cite{ex1}}, label={lst:ex1}]
private AtomicLong counter = new AtomicLong();
public void Increment(long value) {
	this.counter.Add(value);
}
\end{lstlisting}

For one, data sparsity can lead to bad invariants.
Consider the method \lstinline{Increment} in Listing~\ref{lst:ex1} (shown with the associate field for reference); one
inferred pre-condition is \lstinline{this.counter != null}, which is quite appropriate.
However, Daikon also inferred a pre\hyp{}condition \lstinline{this.counter.value == 0}, which is clearly an artifact of
the test suite:
\lstinline{Increment} is only called once per \lstinline{AtomicLong} instance (despite being used many
times in the whole test suite).
Consequently, Daikon also infers \lstinline{this.counter.value == value} as a post-condition,
adding another false positive.

\begin{lstlisting}[language={[Sharp]C}, caption={Method example from Mongo-C\#-Driver \cite{ex2}}, label={lst:ex2}]
public int CompareTo(ElectionId other) {
	if (other == null) return 1;
	return this._id.CompareTo(other._id);
}
\end{lstlisting}

Secondly, we found that the discovery of valid invariants alone was often not the most appropriate criterion. Consider the method in Listing \ref{lst:ex2}, which implements a conventional comparator method (for sorting object types). Here, the obvious (and only reasonable) pre-condition is \lstinline{this._id != null}, and Daikon does indeed infer this. Unfortunately, it also infers eight other related invariants, such as \lstinline{this._id._increment == this._id._timestamp}. Although all of these fields are associated with this method (since \lstinline{CompareTo} is invoked on \lstinline{this._id}), presenting these is in no way helpful to a programmer. This is neither a matter of validity nor of relatedness to the method; Daikon often suggested method-related, valid invariants that were by no means useful. A common example was a post-condition that asserts that a value had not changed, when this was clearly true (produced for both \lstinline{_id} values here).

\begin{lstlisting}[language={[Sharp]C}, caption={Method example from DotLiquid \cite{ex3}}, label={lst:ex3}]
public static void RegisterSafeType(
    Type type, string[] allowedMembers) {
	RegisterSafeType(type, x =>
		 new DropProxy(x, allowedMembers));
}
\end{lstlisting}

Contrary to classical examples of invariants on isolated code (e.g.
stand-alone library classes \cite{polikarpova2009comparative}), our analysis showed that pre- and post-conditions in
real-world methods are often informed by complicated interactions with their various contexts.
However, capturing this complexity via automatic tools appears out of reach.
The code in Listing \ref{lst:ex3} invokes an overloaded method.
Daikon generates the following pre-condition:
\vspace{-.1cm}
\begin{lstlisting}[language={[Sharp]C}]
Contract.ForAll(allowedMembers, x =>
	 x.ToString().OneOf("Name", "ToString"))
\end{lstlisting}
\vspace{-.1cm}
meaning that every element in \lstinline{allowedMembers} is either ``\lstinline{Name}'' or ``\lstinline{ToString}''. Such elaborate invariants are not uncommon for Daikon; we observed many instances of such ``OneOf'' checks (with and without ``ForAll''), all invalid, as well as checks relating elements in arrays with other values or arrays. These invariants constitute Daikon's attempts at inferring complex relationships within the code but are nearly always incorrect.
Kim \& Peterson bring up a similar issue; they find that any semantically meaningful properties of their programs were almost never recognized. Instead, Daikon produces a combinatorially large number of invariants on the relationships across values in the code, with little validity \cite{kimevaluation}.

Even in these simple methods, a complex picture of invariants emerged that highlights several major challenges to useful invariant inference: \begin{inparaenum}\item understanding which invariants are likely to generalize, \item recognizing which invariants are relevant, and \item inferring semantically appropriate invariants\end{inparaenum}.
One might argue that the primary issue in all these cases is that the test suite did not capture sufficiently diverse uses of these methods, but in modern practice this will always be the case: our programs are all well-tested; the problem is that program state on such large programs is far too complex to adequately capture based on just executions. We argue that the heart of the issue is that Daikon infers its invariants on \emph{trace data} alone and has no mechanism in place to validate whether these invariants make sense given the \emph{source code}. We propose to use deep neural networks to extract semantic insights from source code in order to complement invariants extracted from trace data. Here we specifically focus on the first challenge above: judging the validity of a method's pre- and post-condition from its syntax. Our models and results show great promise at this task, suggesting that the other challenges may also be in reach. More generally, our work demonstrates that a great deal of run-time information can be found in the code itself and can statically be extracted by sufficiently intelligent models.

\section{Approach}
\label{sec:approach}
\begin{figure}
\includegraphics[trim=0 280 300 0, width=\linewidth]{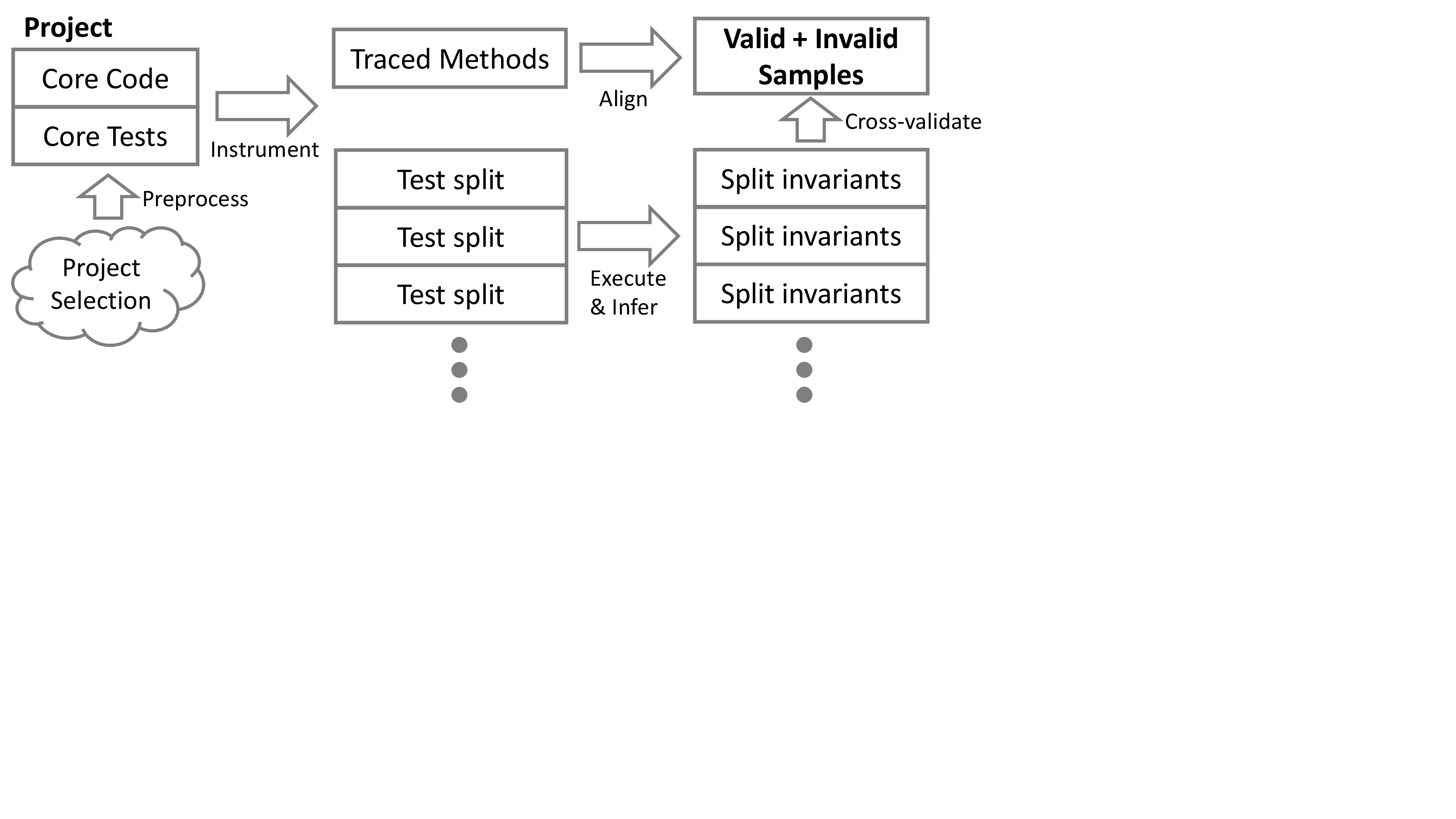}
\caption{Schematic overview of our approach to mining valid and invalid method-aligned invariants}
\label{fig:setup}
\end{figure}

Our goal is to collect a large corpus of methods with their associated pre- and post-conditions, including both valid and invalid examples. This required a number of steps, each with substantial hurdles to overcome, which are shown in \Cref{fig:setup} and discussed in this section.
In addition, we document the process by which we manually curated a ``golden'' dataset and the metrics used in our study.

\subsection{Mining Invariants}
\label{sec:approach:mining}
We aim to rank pre- and post-conditions produced by tools such as Daikon based on their estimated validity.
Any model that can learn to do so requires a corpus of methods aligned with good examples of both valid and invalid
invariants.
The latter category poses a challenge: tools like Daikon produce invariants by running a program's test suite, producing
only statements that are valid across this test suite.
We propose a simple solution based on cross-validation: we sample a subset of test cases and infer invariants on trace
data obtained by running this subset.
We repeat this process many times and compare the invariants obtained on each random subset; invariants that hold on all
of these must hold on the full test suite as well, whereas invariants that hold on only some of the test sets are
invalid.
Crucially, these invalid invariants are nonetheless plausible, as they must have had sufficient support to occur within
some test sets.
This section details the specifics of our data gathering process.

\subsubsection{Data Source}
Since we rely on executing unit tests for our pre- and post-condition mining, we require a sufficiently large and
diverse corpus of runnable projects with adequate test coverage.
An appropriate dataset was mined in recent work on loop idioms, which collected a corpus of 11 C\# projects with diverse functions, for which they were able to run the test suite \cite{8355713}.
We cloned the corresponding Git repositories and reverted them to the same commit mentioned in that work.
Each of these projects (or, in C\# terms, ``solutions'') is comprised of several sub-projects, generally including at
least one ``core'' module and one primary test project with tests for the core functionality; we focus on these two.

As part of our data extraction, we need to be able to execute a random subset of the test suite repeatedly.
Common test suite toolkits do not support such functionality;
instead, we implemented a crawler that extracts test cases and generates code to invoke a subset of them (including any necessary setup/teardown code).
Not all projects in our dataset were amenable to this process; in total, we were able to instrument eight projects in
this way, which are detailed in \Cref{tab:projects}.
Since we only focus on core project tests, manually extract test invocations, and execute a reduced number of projects,
our dataset does not span as many test cases as prior work's \cite{8355713}.
Nevertheless, we were able to extract ca. 43K pre- and 51K post-conditions, each aligned with some 2.5K methods that
average around 100 tokens, producing a dataset that is adequately large for training a deep neural network.

\begin{table}[t]
\caption{The projects used in this study and the characteristics of the invariant-related data extracted from them. We separately extract Pre- and Post-conditions and list the number of methods (\#Mth) and Invariants (\#Inv) collected for each of these.}
\begin{tabular}{ r | l | l | l | l | l }
\multicolumn{2}{c}{} & \multicolumn{2}{| l}{\textbf{Pre-Cond.}} & \multicolumn{2}{| l}{\textbf{Post-Cond.}} \\
\textbf{Project} & \textbf{Tests} & \textbf{\#Mth} & \textbf{\#Inv} & \textbf{\#Mth} & \textbf{\#Inv} \\
\hline
\emph{DotLiquid}			& 356 & 136 & 1,046 & 130 & 1,277 \\
\emph{LibGit2Sharp}		& 745 & 572 & 12,709 & 406 & 13,222 \\
\emph{Logging-Log4Net}	& 184 & 395 & 7,120 & 404 & 7,841 \\
\emph{Lucene.Net}		& 2,585 & 159 & 4,379 & 152 & 6,864 \\
\emph{MathNet-Numerics}	& 2,005 & 880 & 9,637 & 905 & 11,817 \\
\emph{Metrics.Net}		& 168 & 144 & 2,598 & 160 & 3,532 \\
\emph{Mongo-C\#-Driver}	& 1,083 & 187 & 4,211 & 197 & 5,252 \\
\emph{Nustache}			& 146 & 60 & 1,611 & 63 & 1,781 \\
\hline
\textbf{Total} & \textbf{7,272} & \textbf{2,533} & \textbf{43,311} & \textbf{2,417} & \textbf{51,586}
\end{tabular}
\label{tab:projects}
\end{table}

\subsubsection{Trace Extraction}
To extract trace data from C\# DLL executions, we forked and extended Celeriac, a custom Daikon front-end for C\#.
It instruments every method at its entrance and (each possible) exit by
storing the value of every variable in scope, either in terms of an object address (if applicable), or the raw data
value (for primitives, arrays, etc.).
For each project, this front-end was used to \begin{inparaenum}\item instrument the core-project's DLL on which the test project relied to trace all its data, and \item run the test executable many times, writing trace files to a
designated location.\end{inparaenum}\footnote{Some test methods got traced as well; we simply discarded these.}
To reduce the tracing overhead on very frequently used methods, we switch to \emph{method call sampling} on them, which uses exponential back-off in the tracing frequency for each method.
Specifically, the first 10 calls are always traced, followed by sampling of calls $20, 30, \shortdots, 100, 200,
\shortdots, 1000, 2000,$ etc.

\subsubsection{Extracting Invariants}
Our next goal is to extract invariants that are ``scored'' based on their cross-validity between traces. Here, we first (separately) inferred method pre- and post-conditions on each trace file and then cross-validated each with the other trace files. To illustrate how this works, consider a method \lstinline{abs(val)} that returns the absolute value of its parameter, and an appropriate post-condition for this method: \lstinline{val >= 0}. If this method call is observed in 25 traces and the invariant \lstinline{val >= 0} is supported by each of them, it is considered ``valid''. As a consequence, any invariant marked that was supported by the full test suite will also be marked as ``valid'' in our cross-validation.\footnote{It is a remote possibility that some test case is not selected in any split, but the odds of this are less than three in 100,000.} What if, due to limitations of the test suite, five of the trace files instead produced the invariant \lstinline{val >= 1} (i.e. it was never invoked with \lstinline{val = 0})? This invariant does support the weaker post-condition \lstinline{val >= 0}, but not vice versa; we would thus assign it a validity score of 0.2 (5/25) and the corresponding label ``invalid''.

This ``invalid'' post-condition is actually a very useful training sample to our model, for an important reason: invalid
examples such as this one provide examples of plausible yet false invariants with respect to the project's test suite.
This matters because many of the invariants marked as ``valid'' in our method are not actually so; they are only
``valid'' modulo the quality of the project's test suite. We cannot train a model to learn to judge these as valid
without good counter-examples, because it would become far too lenient and not generalize well; randomly generated
invalid invariants thus won't do. In contrast, we argue that \emph{\bfseries extracting examples of invalid statements as we do simulates the inadequacy of the test suite}. This should allow a good model to extrapolate to recognize invariants that are marked as valid but are probably not (as we shall see in \Cref{sec:goldeneval}) and is a contribution of our work.

We instrument each project and run 10\% of its test cases a total of 100 times, storing a trace file for each run.
Ideally, at this point we would run Daikon on every trace file and cross-validate its invariants; it supports an
``InvariantChecker'' mode that cross-references invariants with a collection of trace files, appropriate for our needs.
In practice, however, running Daikon this often was prohibitively slow on larger projects (e.g. one run on Lucene
took over 5 hours), and, more importantly, Daikon's most complex (and error-prone) invariants were also most likely to
not be marked ``invalid'' simply due to data sparsity.
Rather than rely on extensive filtering of Daikon's invariants, we implemented a simple engine that infers a subset of
its language of invariants that we found most likely to be valid in our manual annotation.
It infers the following invariants:

\begin{itemize}[topsep=3pt,itemsep=0ex]
\item Nullity checks on objects.
\item String equality checks
\item Numerical constraints: equality and upper/lower-bounds, relative only to common numbers: $\{-1, 0, 1\}$, $2^n$ and $2^n - 1$ for $n \geq 4$ and $10^n$ for $n \geq 2$.
\item For arrays: all of the above if they hold true for every element in an array (and nullity also if any element is null) and numerical constraints on its length (same values as above).
\item Relational hypotheses: equality between objects, whether an object is contained in an array and relations between two numerical values (where one may be an array length).
\end{itemize}
Our tool is able to rapidly process trace files, both creating invariants on each split and globally cross-validated invariants (for methods that occurred in at least 10 splits).

\subsection{Golden Data}
\label{sec:golden}
We also produce a trace file based on executing every test in a project's test suite. This trace file can be provided to Daikon for it to infer invariants on. Due to the aged nature of the front-end we used, we had to recompile Daikon from source with added error handling in order to process most of the resulting trace files; only Lucene could not be run at all.\footnote{Typical issues included method traces not matching their signature; Daikon did not provide an option to skip such cases.} From the resulting invariants, we selected five methods at random that had at least two pre- and post-conditions for each project, yielding 217 pre-conditions and 282 post-conditions.

We manually labeled these invariants by inspecting the corresponding code. Our study design was as follows:
\paragraph{Raters:} We had access to four raters, each of whom had at least five years of experience in software development and familiarity with C\#.\footnote{One rater had limited experience with C\# but extensive experience with Java and C, which are closely related.} Each method was randomly assigned to two of our raters to separately assess all its pre- and post-conditions.
\paragraph{Inspection:} Every method was considered in the context of its documentation (if present) and any reasonably related code, including fields of the surrounding class, signature of related types, methods that invoked it, and methods that it itself invoked. Each invariant was annotated as either ``valid'', ``invalid'', or ``irrelevant'' according to the following criteria.
\paragraph{Irrelevant:} Any invariant that was not in any way related to any variable that played a role in the method (field, parameter or local) was marked as ``irrelevant''. This includes many class invariants that were unaffected by the method, but not e.g. invariants on fields of objects that were used in the method.
\paragraph{Invalid:} Any relevant invariant was marked as ``invalid'' only if there was convincing evidence that it did not hold. This includes many cases of test suite sparsity (some shown in \Cref{sec:motivation}), such as arrays only ever containing one element, or a parameter object never being set to null despite this being clearly allowed,\footnote{i.e., it is either documented as such, marked as nullable, or checked for nullity in the method body} as well as overly complex invariants that were incorrect based on inspection of the semantics of the method and class.
\paragraph{Valid:} The remaining valid invariants are likely to be a slight overestimate, but contain many reasonable checks. Virtually all of these were very simple, such as nullity checks and comparisons with simple numerals. We additionally marked many ($\sim$17\%) of these as ``not useful,'' when the invariant stated a property that was in no way relevant to the semantics of the method. An example of such an invariant was given in \Cref{sec:motivation}; nearly all of these were post-conditions asserting that a value that was not affected by the method had indeed not changed. As this is a more subjective call, we do not include this judgement in our analysis.
\paragraph{Conflicts:} After every rater had made their pass, some 25 invariants were given conflicting ratings. Many of these were found to be due to the complicated nature of identifying an invariant as ``(ir)relevant''. In response, we adopted the aforementioned guideline regarding ``irrelevant'' invariants, i.e. that an invariant is relevant if it contains any reference to a value used in any way in the method (including, e.g. to a field of an object used in the method), thus preferring not to assign ``irrelevant'' unless absolutely appropriate. This resolved most of the conflicts and the guidelines were improved to incorporate those decisions. The remaining conflicts were discussed by their two raters in order to establish whether an objective conclusion could be reached (e.g. if one of the raters had overlooked a null-check). This left four unresolved conflicts, which were omitted from our dataset.

The resulting corpus will be referred to as our ``golden'' data. Our analysis identified less than one in three invariants as valid; most others were false positives, either because they were not relevant at all or because they were invalid. Note that, since we adopted a low threshold for considering invariants ``relevant'', a more stringent analysis would mark some of the ``invalid'' invariants as ``irrelevant'' instead; this does not affect our ``valid'' ratings. Note that we mainly focus on distinguishing between the ``valid'' and ``invalid'' invariants in our analysis (see \Cref{sec:goldeneval}), as our model is not trained to recognize relevancy.

\begin{table}[t]
\caption{Validity statistics of Daikon's inferred invariants on random method sampled from the studied projects.}
\begin{tabular}{ r | l | l | l }
 & Overall	& Pre-conditions	& Post-conditions \\
\hline
Irrelevant & 25.7\% & 23.3\% & 27.5\% \\
Invalid	 & 42.6\% & 50.2\% & 36.8\% \\
Valid	 	 & 31.7\% & 26.5\% & 35.7\% \\
\hline
\end{tabular}
\label{tab:daikon}
\end{table}

\subsection{Metrics}
\label{sec:metrics}
Since our objective is to rank invariants based on their validity, our primary means of evaluation will be the Receiver Operator Characteristic (ROC) curve. The output of our models is the predicted probability of an invariant being valid. When analyzing our results, we can simulate setting a threshold on this probability, above which we include all predictions. Each threshold will yield a different balance between valid and invalid invariants, where ideally high threshold yield a very favorable balance of valid to invalid invariants. The ROC curve quantifies this trade-off; it plots the True Positive Rate (TPR, y-axis) against the False Positive Rate (FPR, x-axis), to produce a curve that ranges from (0, 0) to (1, 1), where random guessing yields a straight line between these points.\footnote{A related curve is the Cost-Effectiveness Curve, which instead plots the proportion of all inspections on the x-axis (so it produces lower AUCs); we evaluated both and found comparable results, so we present the more common metric.} The Area Under Curve (AUC) for the ROC curve is commonly used to assess how much ``lift'' an ROC curve achieves compared to random guessing, which would yield an area of 50\%. We present both this summary metric and the overall curve; see \Cref{fig:cross} for an example.

We investigate two settings for ranking: project-wide and per-method. The latter setting computes the ROC value for ranking invariants for each method that has both valid and invalid invariants and is arguably the most realistic real-world use of our tool. However, since many methods had only invalid invariants, we also produce ROC curves across each full project. This corresponds to a scenario where a developer annotates a project with invariants without focusing on any particular method.

\section{Model Architecture}
\label{sec:architecture}
A complete analysis of invariant validity requires inspecting the entire code context of the method under investigation.
However, we argue, relying on prior work~\cite{allamanis2018learning}, that such semantic analysis can often be
circumvented via inspecting the syntactic patterns of the method's identifiers, control flow, and signature.
This is because source code is \emph{dual-medium} in nature: developers write it to communicate some intent
simultaneously to the machine (as formal instructions) and to fellow humans (as natural language).
We hypothesize that a model imbued with natural-language processing capabilities is able to extrapolate these syntactic
patterns to accurately judge semantic validity of many code properties, including invariants.
Thus, we restrict ourselves to the method source code as the sole source of information, providing an already accurate
lower bound on the potential ranking performance.
Other promising data sources may be studied in future work (\Cref{sec:extensions}).

Capturing meaningful semantic information even from just a method requires a powerful model.
The average method in our corpus contained over 100 AST nodes (including leafs), with many several times larger.
SE studies commonly use Recurrent Neural Networks (RNNs), which digest a sequence of lexical tokens from left to right
(and possibly right vice versa) to emit a representation for each token.
Although these models can learn useful representations, they often struggle to aggregate data across long distances that
are very common in code \cite{allamanis2018learning, hellendoorn2017deep}.
More importantly, they cannot represent relational and structural properties of source code, required to infer its
semantics accurately.

Graph-based neural networks were proposed as an alternative.
These networks represent source code as a directed graph with different domain-specific kinds of edges, including edges
between lexically adjacent tokens (which RNNs use), but also syntax tree edges (between parent and child nodes) and
long-distance edges between related tokens (e.g. data-flow edges).
By directly modeling these relations, information can flow across long distances directly between semantically relevant
locations, mitigating many of the problems with RNNs.
We adopt this model in our work, specifically focusing on Gated Graph Neural Networks (GGNNs), which were shown to
learn generalizable, compact representations of source code \cite{li2015gated, allamanis2018learning,
brockschmidt2018generative}; we also include RNNs in our ablation analysis.
For a detailed description of this model, see \cite{allamanis2018learning}; we briefly give an overview of its
behavior here.

A GGNN takes as input a graph representation $\mathcal{G}$ of the method, constructed by enriching its AST with
additional kinds of edges between related nodes, such as lexical order edges and data-flow edges.
Every node $v \in \mathcal{G}$ has a \emph{state} $\bm{h}_v \in \mathbb{R}^m$, its vector representation.
These states are initialized with node \emph{embeddings}, learned end-to-end from the nodes' \emph{features}
(e.g. nonterminal kinds, data types, and identifiers for leaf nodes).
Like prior work, we also split compound identifiers by camel-case and underscores and embed such nodes by summing over the embeddings of their subtokens.

The GGNN performs $K$ phases of \emph{message passing} to infer the representations of all nodes used to determine the
network's task output.
In each phase, every node $v$ passes its state to all adjacent nodes as a \emph{message}.
Every edge $\langle v, kind, w \rangle$ transforms the passed message using a global weight matrix, shared for all edges
of the same \emph{kind}.
Finally, every node $w$ aggregates its received messages and updates its own state based on these using a recurrent
cell function,
allowing it to select which part of the received information to store and which of its own state to preserve.
After $K \geq 8$ phases the final representations of each node accurately capture their role in the method, and
thus can be used to determine the task output.

In our approach, the GGNN's inference task is invariant validation for a given $\langle \text{method}, \text{invariant}
\rangle$ pair.
To compute it, we first inject the invariant (pre- or post-condition) into the method as appropriate.
We then perform $K$ message passing phases of the GGNN, after which we aggregate the hidden states belonging to all
nodes in the \emph{invariant}.
The resulting state is passed through a hidden linear layer and then projected onto one dimension (with sigmoid
activation), producing a probability judgement.
This architecture can be trained end-to-end using the data described in \Cref{sec:approach:mining}, as we outline below.

\subsection{Training Setup}
We implemented a highly optimized GGNN for source code in TensorFlow Eager~\cite{tensorflow}, operating on top of method graphs automatically extracted using the C\# compiler interface.
Our implementation can process around 350 methods at the time (in minibatches of up to 40,000 tokens) on a GTX 1080Ti GPU, requiring 3-5 seconds each.
We were able to train ten epochs on the full corpus in around two hours; our full set of experiments completed within a
few days.

For our cross-project setting (our main goal), a model for each project was trained on all other projects. We held-out a small portion of the test project as validation data to track overfitting of our models; we stress that this was only acceptable because this is a cross-validation experiment -- holding out part of the training data would yield a very poor estimate of overfitting (see intra- vs. cross-project results in \Cref{sec:intracross}). In practice, our model's performance tended to saturate on the held-out data within just a few epochs and overfit starting around epoch five, likely because of the high degrees of intra-project duplication. We therefore generally evaluated the models at epoch three, unless stated otherwise.

We trained our model using an Adam optimizer \cite{kingma2014adam} with learning rate of 0.001. A vocabulary was
estimated on the training corpus and words occurring only once there, as well as new words in the test data, were
treated as ``<unknown>'' tokens.
We excluded methods with more than 500 nodes, which constituted a negligible portion of our
data. Finally, ablations of the (twenty) edge types proposed in prior work suggest that just six of these are necessary
for building accurate models (which greatly accelerates training): lexical adjacency edges (next-token \&
previous-token), AST edges (parent \& child) and lexical use edges between occurrences of the same identifier (next-use
\& last-use, which approximate data-flow relatedness).
We reduced our edge types to just these.

We also trained a conventional RNN model; this model was trained to encode the method and invariant separately, aggregate each (by averaging across their tokens) and judge the validity of the invariant by passing the two encodings through a hidden layer and projecting to a probability judgement (like the GGNN's discriminator). We used 300-dimensional embeddings (shared by the method and invariant encoder) and two 500-dimensional bi-directional encoders with GRU units (one for each input); these parameters are in line with RNNs typically used in modeling source code \cite{hellendoorn2017deep, Hellendoorn-DLTI}.

\section{Results}
\label{sec:results}
We start our analysis with the data that we mined at scale using the methods described under \Cref{sec:approach}.
Although this data likely contains a high degree of false positives, it allows us to evaluate whether our models can extrapolate from the training data, both within and across projects, as well as the factors that contribute to
their performance (in an ablation study).
Finally, we revisit our manually annotated data to assess the true validity of our models.

\subsection{Corpus Data}
\label{sec:intracross}
The simplest case for our models should be to train and test on methods from the same project.
This ensures that the identifier vernacular at train and test time is highly similar (which is always beneficial for
deep learners~\cite{hellendoorn2017deep}) and allows the deep learner to profit from any repeated invariants within the
project (which our manual analysis suggests is common).
This setting is not unrealistic: when deployed as a tool, our validator could be trained on part of a project
in order to rank inferred invariants on new methods or ones with low test coverage.
For most practical purposes, however, cross-project performance is more appropriate; ideally a model can be pre-trained
on our corpus and extend well to unseen projects. We evaluate both settings.

\begin{figure}[t]
\includegraphics[trim=0 0 0 0, width=0.495\linewidth]{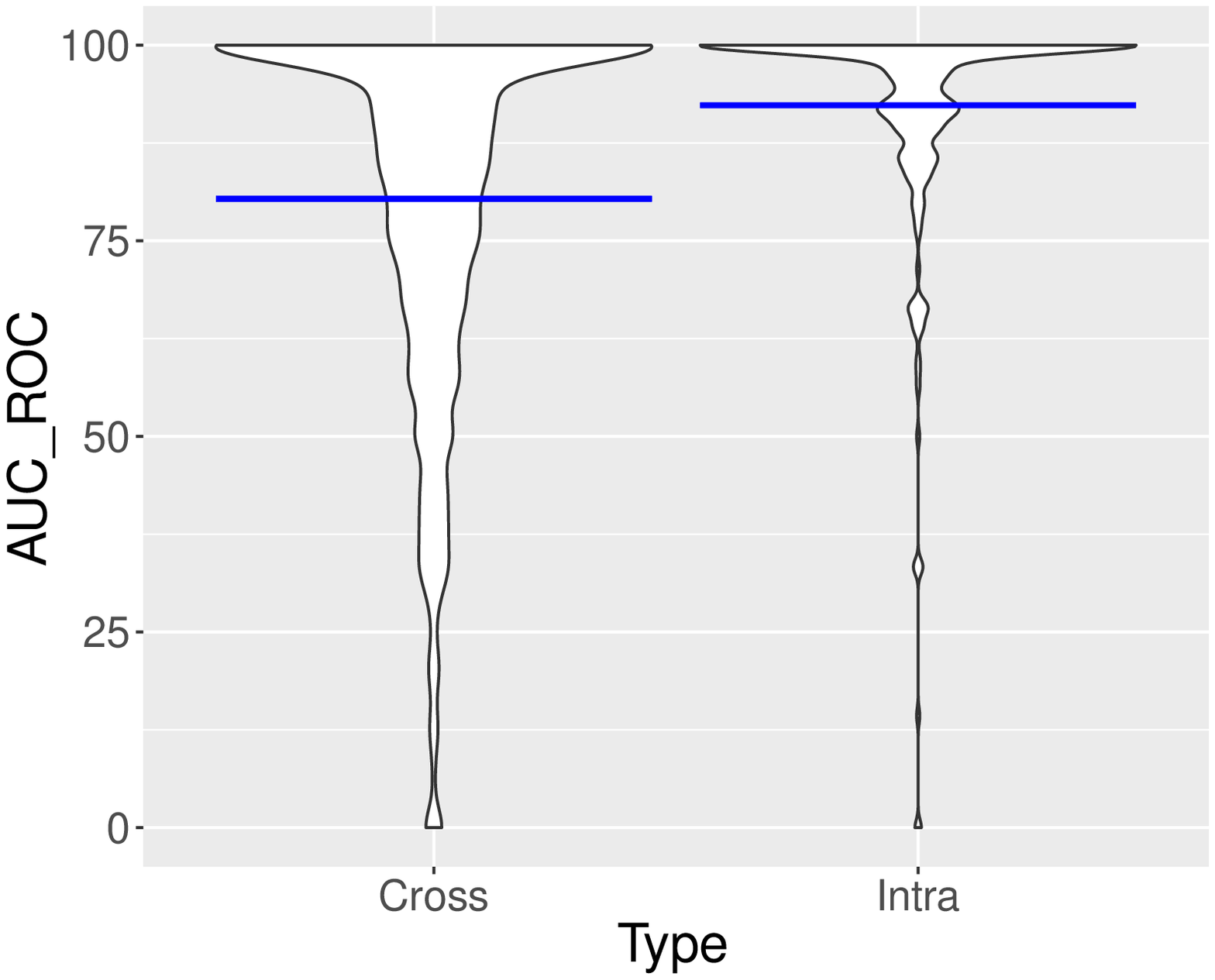}
\includegraphics[trim=0 0 0 0, width=0.495\linewidth]{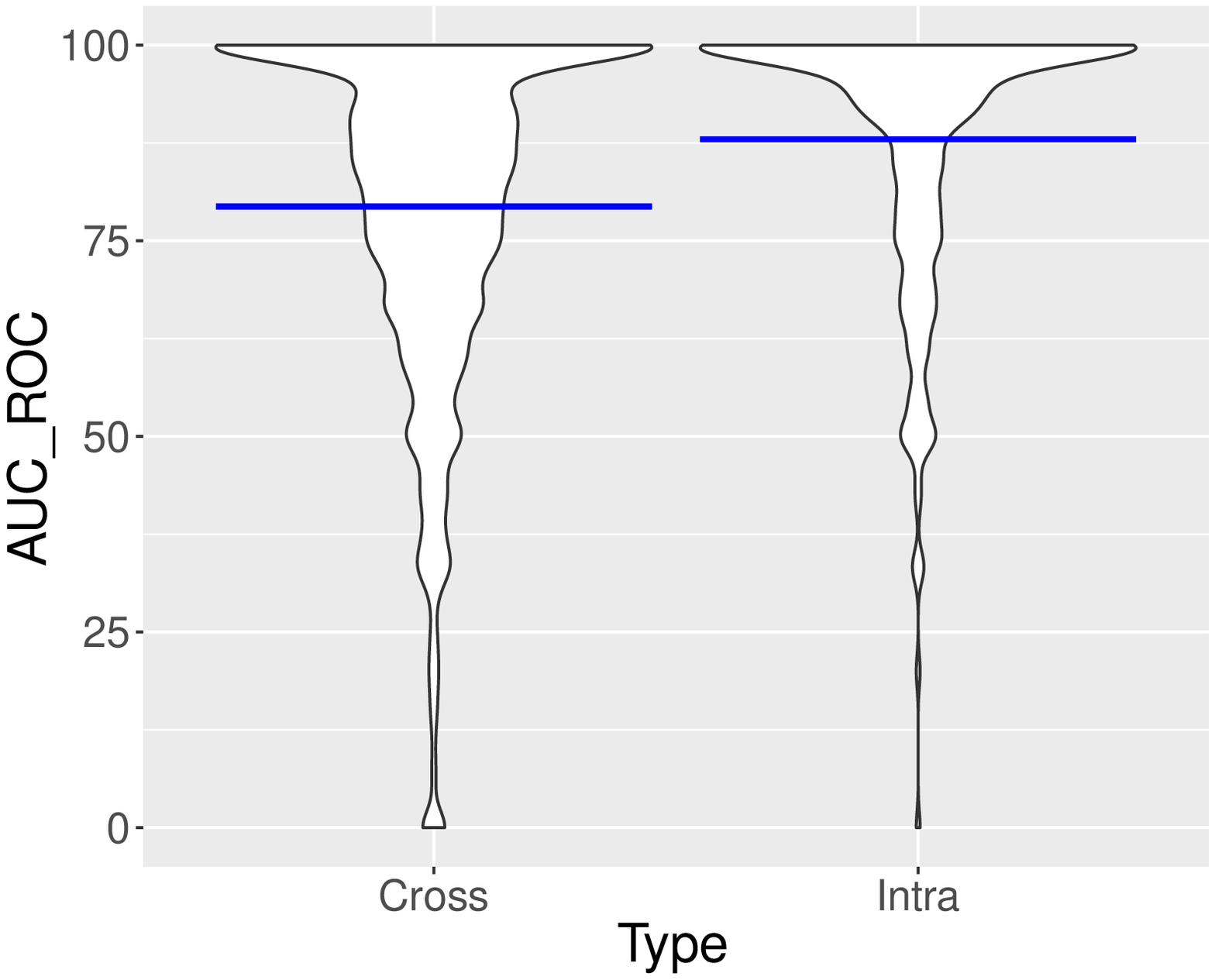}
\caption{Per-method AUC-ROC distribution with their means for intra-project and cross-project models. Left: ranking pre-conditions, right: ranking post-conditions.}
\label{fig:intra-cross}
\end{figure}

We show the per-method results for both intra-project and cross-project invariant ranking in \Cref{fig:intra-cross}. Intra-project predictions are evidently very easy in our dataset; ROC values of over 90\% signal near-perfect predictions. This is not to boast of the quality of our model; it more likely signals a high degree of redundancy in the invariants within each project. Although that might impair the performance of a machine learner, our GGNN's cross-project performance is actually quite good; it even perfectly ranks invariants in a third and a quarter of pre- and post-conditions respectively.
The mean values between pre-conditions (intra: 93.8\%, cross: 76.5\%) and post-conditions (intra: 91.9\%, cross: 76.3\%) were quite similar, though the latter had a wider spread of success. 

\begin{figure*}[t]
\includegraphics[trim=0 50 0 0, width=0.495\linewidth]{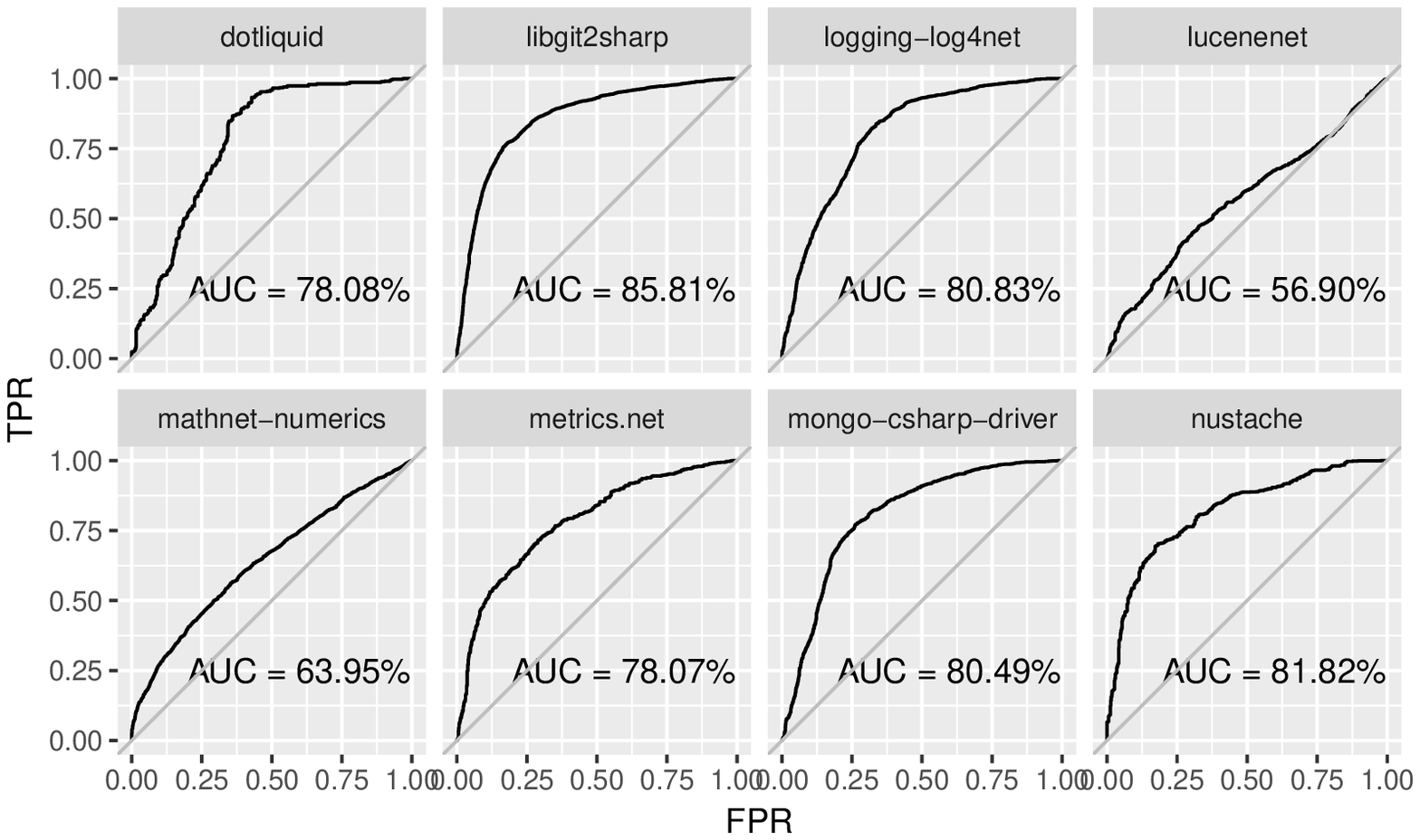}
\includegraphics[trim=0 50 0 0, width=0.495\linewidth]{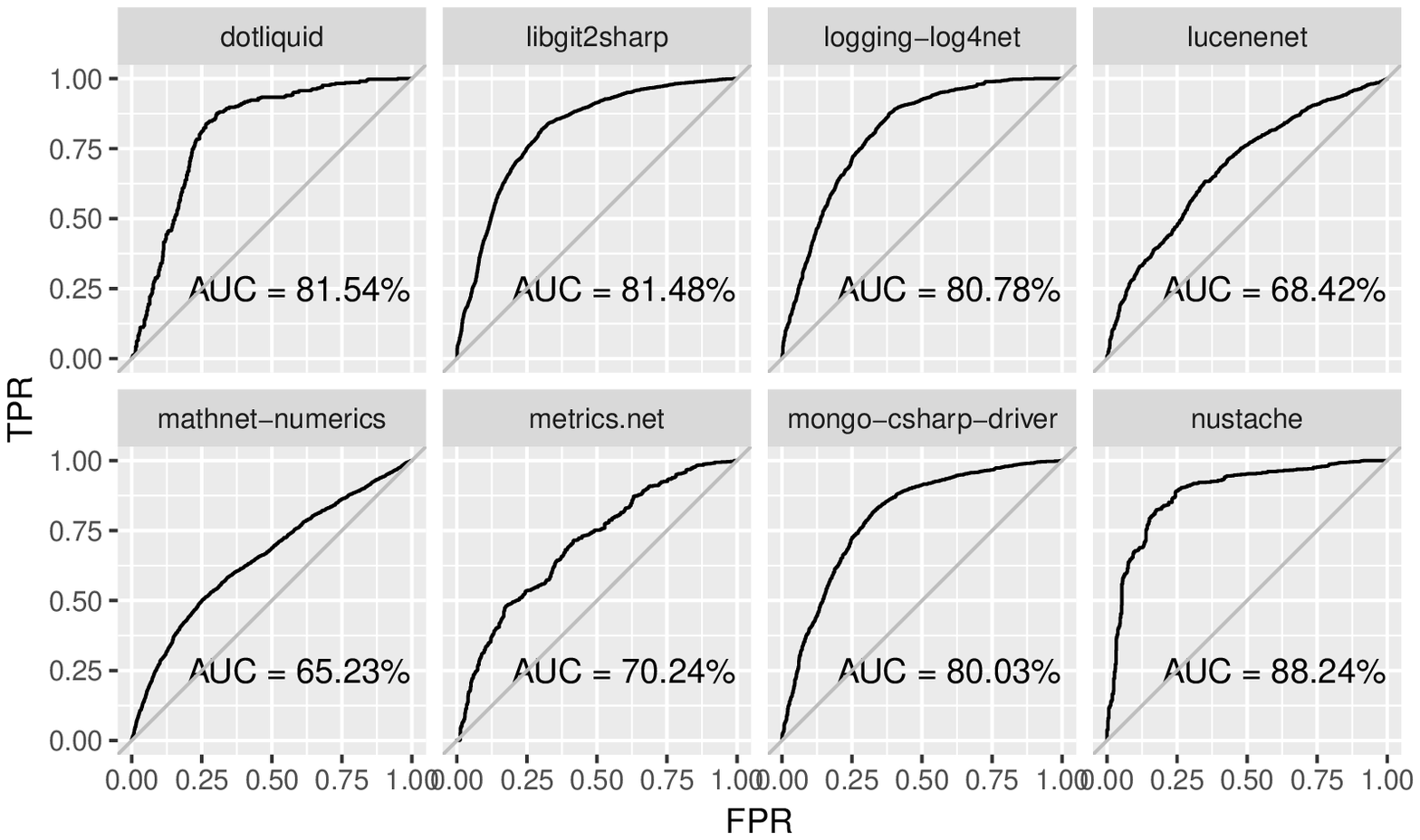}
\caption{Cross-project ROC curves with their AUC values; grey diagonal line shows expected values for random guessing. Left: ranking pre-conditions, right: ranking post-conditions}
\label{fig:cross}
\end{figure*}

A closer look at the per-project performance in terms of ROC curves is given in \Cref{fig:cross}. Here we rank all invariants in a project by their projected score (see \Cref{sec:metrics}). Most projects achieve quite reasonable ROC curves, with AUC values in the range of 75\%-82\%; only Lucene in pre-condition ranking appeared to pose a major challenge to our model. These curves also provide more insight into the practical ranking behavior on these projects; if a developer were to only inspect a small portion of invariants (not unlikely given the large number of inferred statements), the left-most section of each graph shows the trade-off they would experience. In many cases, the ranking performance there is substantially better than on the whole project; especially for Mathnet-Numerics and Metrics.net, the initial section of this curve is very steep, suggesting that the highest ranked invariants are predominantly valid. We can quantify this too: at a 25\% FPR, random ordering would produce an AUC of 3.125\%, but our models exceed 10\% on both conditions, outperforming random ranking by a factor three; at 5\% FPR, they outperform it by more than five times.

\subsection{Ablations}
\label{sec:ablation}
To get more insight into the quality of our model and the factors that contribute to its accuracy, we run several ablation studies. One ablation reduces the complexity of our model: our graph-gated neural network combines several different edge types,
including adjacent tokens, but also syntax tree edges and edges between occurrences of the same identifier. More
commonly, work on modeling code has used Recurrent Neural Networks (RNNs) \cite{White:2015:TDL:2820518.2820559, Hellendoorn-DLTI}, which consider only adjacent tokens. We train such a model as well (see \Cref{sec:architecture}).

Our second ablation is aimed at better understanding whether our GGNN actually captures any semantics from the method body. To evaluate this, we train a GGNN (with the same architecture) that makes a validity judgement just based on the invariant itself (ignoring the method body). If an invariant can be accurately judged based on just its content, this suggests that our GGNN does not capture much additional information from the method body, either because it cannot or because it is often unnecessary. This might not even be surprising: we found many invariants that were only weakly related to the method in our analysis. In a sense, our full model establishes a lower-bound: how much ranking performance can be achieved with just the limited information that the method body provides? This ablation helps us better understand this lower-bound.

This turns out to be an interesting study, for two reasons. First, the RNN model performs a fair bit worse than the GGNN, which is to be expected because due to its lesser capacity. However, it also performs worse than the ``No Context'' model, which does not consider the method at all, despite sharing the same training data and achieving high training accuracy (meaning the model was able to learn well). This strongly suggests that the RNN model learns patterns that poorly generalize. This is not just a matter of overfitting; its test performance tended to peak around the third epoch, when training accuracy was far from convergence. It has been documented before that GGNNs are better at generalizing from training data \cite{allamanis2018learning}, and this finding confirms that result.\footnote{Partially this may be due to GGNNs better using a limited parameter space; our GGNNs use embedding and hidden state dimension of 128, whereas the RNN uses 300 \& 500, simply because it does not learn good representations with lower values.}

This brings us to the second ablation: at face value, the ``No Context'' model performs almost identically to the full GGNNs, losing by just 2-3\% on pre-conditions and 1\% on post-conditions. This would seem rather disappointing; it suggests that the naturalness of a method betrays very little about an invariant's validity that is not already contained in the invariant itself. The small difference in AUC-ROC is significant and may be beneficial to developers,\footnote{The full model especially outperformed it on Lucene, our hardest project, where the no-context model never exceeded random AUC-ROC} but would appear to make it hardly worthwhile to train the full model. Fortunately, however, this result does not accurately reflect reality. The perceptive reader may have noticed the final column, which does paint a stark contrast; we will discuss this next.

\begin{table}[t]
\caption{Performance of ablations of our model compared to the full model, expressed in AUC-ROC values for pre- and post-conditions, both for ranking invariants across the whole project (averaged over projects) and for ranking within every method that has at least one valid and invalid invariant (averaged across all methods).}
\begin{tabular}{ r | l | l | l | l || l }
 &			\multicolumn{2}{l |}{Pre-conditions} & \multicolumn{2}{l ||}{Post-conditions} \\
\textbf{Model} & \emph{Project} & \emph{Method} & \emph{Project} & \emph{Method} & \emph{Golden} \\
\hline
Full Model		& 76.5\% & 80.4\% & 76.3\% & 77.9\% & 83.0\% \\
RNN			& 72.5\% & 76.5\% & 68.6\% & 73.4\% & 60.0\% \\
No Context 	& 74.0\% & 77.9\% & 75.1\% & 77.1\% & 60.9\% \\
\hline
\end{tabular}
\label{tab:ablation}
\end{table}

\subsection{Golden data}
\label{sec:goldeneval}
All of the aforementioned results were based on the automatically mined corpus, which we know contains many false positives still. Training with such flawed data was a deliberate decision: our goal was not to learn to replicate the labeling, but to recognize when plausible invariants are \emph{likely} to be invalid. This is why our invalid invariants are mined by holding out test suites; we argue that this makes the model more likely to infer which invariants are valid just due to sparsity of the test suite. If our deep learner is adequate, it should be able to recover this signal despite the overwhelming amounts of noise of invalid invariants being presented as valid at training time. The end-goal of our learner is thus not to score well on our automatically mined test data (although it certainly does), but to perform well in a real-world evaluation. To accurately assess its performance at this task, we turn to our golden data.

We use our deep learners to score all invariants that we manually annotated, removing those that were not in the scope of our invariant extractor (we omitted learning from certain invariant types because they were highly unlikely to be valid, see \Cref{sec:approach}). The resulting set contains 37\% irrelevant, 26\% invalid and 37\% valid invariants, which resembles to our overall data, but with substantially fewer invalid cases (due to our filtering). All invariants labeled ``valid'' in our manual annotations were also valid in our training data (as should be the case), but false positives abounded: nearly 30\% of the invariants that we manually labeled ``invalid'' were nonetheless presented as valid in our test data. The invariants labeled ``irrelevant'' were all included in the corpus as valid. Although this may be an accurate assessment, these were entirely spurious and our model was not trained to recognize relevancy, so we focus only on distinguishing ``valid'' from ``invalid'' invariants.

The final column in \Cref{tab:ablation} shows the AUC-ROC performance of our various models at ranking the invariants in this dataset, where the goal is to rank ``valid'' invariants above and ``invalid'' invariants. We focus just on ranking the full annotated set here, since there are too few methods to get reliable per-method ROCs. The difference in performance now becomes substantial; the context-free model barely scores better than random, whereas the full GGNN achieves quite good performance. The RNN model does not perform any better either, offering strong evidence that the complexity of the GGNN architecture is necessary to capture the semantic cues required for disentangling valid and invalid invariants.

\begin{lstlisting}[language={[Sharp]C}, caption={Method example from Logging-Log4Net \cite{ex4}}, label={lst:ex4}]
protected override void WriteHeader() {
	if (m_stream != null) {
	...
}
\end{lstlisting}

Listing \ref{lst:ex4} shows a case where the GGNN was able to extract semantically useful information from a method. In this method, the first statement evaluates the nullity of the \lstinline{m_stream} field, implying that the inferred pre-condition \lstinline{m_stream != null} is certainly invalid and should not be ranked highly. The no-context model does not have access to this information and assumes the typical validity of a nullity check: mostly valid, ranking this in the top 20\% of validity scores (the RNN makes a similar call). The full GGNN, on the other hand, ranks it in the bottom half of its judgements, well beneath almost all valid invariants. Despite having seen many contradictory examples at training time (and indeed, assigning it a fairly high probability), it has learned to recognize that there is some doubt about its validity given the nullity-check in the method body (it ranked many valid nullity checks far higher). In doing so, it has accomplished the goal of our training setup.

\subsection{Data Availability}
\label{sec:package}
In the interest of open science, we upload a package of our data (manual annotations and mined graph files), code (GGNN and RNN implementations), and output logs (intra-project performance of GGNN, cross-project performance for all models); available at [\url{https://doi.org/10.5281/zenodo.2574189}]. We will turn this into a de-anonymized, full-fledged replication package when anonymity is lifted.

\section{Discussion}
In \Cref{sec:motivation,sec:approach,sec:architecture,sec:results}, we discussed each of the four contributions of this paper: \begin{inparaenum}\item a manually annotated dataset of Daikon invariants on complex, real programs, \item an approach to mining plausible valid and invalid invariants at scale, \item a novel neural architecture that is able to assess the validity of method pre- and post-conditions based on the method, and \item results establishing that our model is able to train well on our training corpus and uniquely generalizes to manual annotations. \end{inparaenum} Taking a step back, we now discuss the implications, validity and possible extensions of this work.

\subsection{Implications}
The immediate goal of our model was to rank invariants that are proposed by a tool such as Daikon (or our subset thereof). Our results indicate that a developer who uses such a tool to infer method pre- and post-conditions would be well-served by first using our tool to prioritize the output and find invariants that are most likely to be valid. This can substantially reduce the proportion of false positives that are presented to the developer.

More broadly, our work is one in a series of studies that aim to learn to extract program properties, those which cannot be easily or even feasibly be extracted by conventional static analyzers (such as compilers), directly from its
source code by exploiting its naturalness.
This includes work on inferring types \cite{raychev2015predicting, Hellendoorn-DLTI}, learning naming conventions \cite{Allamanis:2014:LNC:2635868.2635883, allamanis2015suggesting} and identifying fault-prone regions \cite{ray2016naturalness, allamanis2018learning}. Like related work on loop invariants (see \Cref{sec:related}), our work focuses on properties that describe the \emph{behavior of the code}, a highly non-trivial artifact of its syntax, which even professional developers struggle to accurately assess. Unlike that body of work, we cannot rely on an SMT solver in the process; our models are stressed to capture a meaningful semantic signal from just the source code. We accomplished this by using a combination of a powerful machine learner and an approach to mining data at scale that is noisy in a way that actually encouraged our model to generalize. These innovations help advance our ability to learn program properties.

Finally, deep learning has especially boosted this field in recent years by providing a family of models that can learn to both achieve state-of-the-art performance in various tasks and learn powerful representations of source code at the same time. The latter is arguably scientifically most interesting: much like how latent representations in computational linguistics and computer vision have proven to capture surprisingly rich semantic features in their domains, even from unlabeled data, representations of source code have the potential to capture much of the semantic insights that developers embed in their programs through identifiers, common idioms and architecture. As such, models that can successfully address complex tasks such as ours likely learn generalizable and worthwhile representations of the programs they study in the process. Our model's success at generalizing from our training data to real annotations, despite their very different characteristics, is evidence of that.

\subsection{Extensions}
\label{sec:extensions}
An immediate extension of our work is to consider more context: comments, the surrounding class, and any invoked and invoking methods (even in other classes) can be highly valuable sources of information. In many cases, the human annotators were not able to confidently infer the validity of the invariant from the method body alone (though it virtually always provides some information). In addition, many of our studied projects were well-documented; although this cannot always be relied on, studying this documentation with deep learning can be highly beneficial \cite{Pradel:2018:DLA:3288538.3276517,nl2type}.

Further study of the appropriate models for our task may be beneficial. We considered several simpler ablations that were less successful, but more complex (or entirely different) architectures may perform better than ours. In addition, it may be possible to integrate trace data values directly into the model; although this slightly changes the task from ours, it may not be unrealistic to already have some tests in place for a method that a developer wants to annotate. Such an extension could provide a valuable step forward in modeling the state of programs directly \cite{wang2018dynamic}.

\subsection{Threats}
Our work required a substantial number of design decisions and solutions, which are important to restate in relation to both its internal and external validity.

We required a set of executable C\# projects, for which we relied on a previously collected dataset. Although we cannot make certain claims regarding generalization to other languages and projects, our corpus was deliberately selected from projects with diverse goals and implementations by prior work. In addition, C\# is a popular language that closely resembles other languages that are often used to build large systems (such as Java, Kotlin and C). However, generalization to more script-like languages (such as Python and JavaScript or TypeScript) is not a given and deserves further investigation.

We instrumented our projects to generate our own test execution harness, causing us to work with a reduced set of projects and test cases. This was hard to avoid given our parameters, but future work may investigate the efficacy of more light-weight approaches (e.g. randomly dropping method traces from one large trace file). We were still able to work with nearly one thousand test cases per project, which produced many gigabytes of trace data. Our instrumentation also required adding improved error-recovery to both Daikon and its C\# front-end (Celeriac), which may introduce some missing values in our trace data and invariants. In practice, however, these errors were rare, occurring only on large trace files.

Our manual annotation carries the risk of errors in human judgement, which we aimed to mitigate by having two annotators per invariant, as well as strict guidelines that allowed for invariants to be marked as ``irrelevant'' besides the usual categories of ``valid'' and ``invalid''. Nonetheless, annotating pre- and post-conditions on large real systems was a complicated assignment and often required extensive investigation of a method's context; we have documented evidence for all of our more complex decisions in our replication package (see \Cref{sec:package}, and\Cref{sec:evaldaikon} for further discussion).

Our ablation analysis considers two simpler alternatives to our full model: a bi-directional RNN and an invariants-only
validator. We designed our own ablations to address (and dismiss) various simpler explanations for our model's efficacy;
to the best of our knowledge, no prior work has addressed this task, so no further baselines were included. Our model
generalized substantially better than either of these, even though (bi-directional) RNNs are powerful models with
many applications to software engineering tasks \cite{Hellendoorn-DLTI,White:2015:TDL:2820518.2820559}.

\section{Related Work}
\label{sec:related}
Our work occupies addresses an unusual topic in aiming specifically at learning method pre- and post-conditions. Although we are not aware of any work that tackles this same issue, there is a fair body of work that studies the learning of loop invariants. In this setting, tools can rely on an SMT solver provided the loop and state-space is small enough. Furthermore, we are not the first to analyze Daikon's output; we discuss both these families of related work here.

\subsection{Learning (Loop) Invariants}
Sharma \etal describe a data-driven approach to finding algebraic (polynomial) invariants using a ``guess and check'' approach~\cite{sharma2013data}: they ``guess'' polynomial invariants by solving a system of linear equations over all possible monomials up to a fixed degree across traced values of variables (one row per observed state vector). They then ``check" these using an SMT solver; any counterexamples the solver finds are used to create another
test input, after which the process is re-run. This approach is subject to the scaling limitations of the SMT solver, as well as the combinatorial growth of matrices with monomial degree and state vector dimension.

A related paper describes the use of PAC-learning to learn integer loop invariants on small single-procedure programs with one loop~\cite{sharma2013verification}. The PAC learner induces invariants as geometric constraints on a vector of loop variables. Labeled examples are obtained from passing test cases that satisfy a post-condition (``good'') and by sampling satisfying values of a weakest pre-condition that would force a post-condition to fail after one iteration of the loop (``bad"). This approach appears to work only on small programs with specified post-conditions (in order to generate bad test cases). On bigger programs, finding such a weakest precondition is not scalable. Our approach learns a compact representation of the program, relevant to the task, allowing it to better scale than both these approaches.

Padhi \etal learn pre-conditions and loop invariants as boolean combinations of a set of arithmetic conditions (``features") \cite{padhi2016data}. They improve upon previous approaches (e.g., ~\cite{garg2016learning}) which could learn such combinations, given an adequate set of features, whereas Padhi \etal can synthesize new features. Their synthesis approach essentially generates and tests all features up to a size, searching for a new feature that can distinguish between passing and failing inputs, thus improving the expressive power of possible feature combinations. This approach requires passing and failing tests and is agnostic with respect to the program structure.
Pham \etal\cite{pham2017assertion} also describe an approach that is agnostic to the structure of the original program, which uses a fixed set of feature templates over state vectors, together with an SVM-based approach, to learn linear inequalities that classify passing and failing state vectors. In addition, they modify the original program to inject artificial states near decision boundaries so as to create new labeled examples, as a form of active learning. The paper does not show that this approach is sound. Although it is one of the few works to address learning pre-conditions, it requires post-conditions to be in place, as well as labeled traces corresponding to passing and failing tests.

Si \etal propose a deep reinforcement learning approach to creating loop invariants via stochastic synthesis \cite{si2018learning}. Ultimately, an SMT solver (Z3) is used to provide (automated) supervision. The approach incorporates a learned GGNN to create representations of a semantic abstraction of a given program, together with an RL approach to learn a policy that directs the synthesis of a loop invariant. The RL reward mechanism finesses the sparsity of the eventual reward (the final validity of the invariant, as labeled by Z3) by creating intermediate rewards; these count and normalize the proportion of counterexamples produced by Z3 by candidate invariants as the generation proceeds. This normalized count is supplemented by another intermediate reward (provided before invariant generation is complete) that is designed to reject ``meaningless" and ``trivial" predicates such \emph{e == e} or \emph{e $<$  e}. Although we do not focus on generating invariants (and cannot rely on an SMT solver), the approach proposed in this paper is quite complementary to ours, suggesting a possible symbiosis (e.g. to tackle the third challenge in \Cref{sec:motivation}: generating semantically meaningful invariants).

Brockschmidt \etal induce invariants over data structures for shape analysis, using a similar approach of generating invariants (in separation logic) production using a learned neural network \cite{brockschmidt2017learning}. The production selection is based on hand-engineered features over the data-structure graphs. Rather than using SMT solvers (with RL) for supervision, they use data produced from test runs. This approach is program agnostic and may thus benefit from insights in our paper, despite considering a very different class of invariants.

\subsection{Evaluating Daikon}
\label{sec:evaldaikon}
Daikon has become a landmark tool in the academic community \cite{ernst2007daikon}. Accordingly, it is involved in much related work; it is often used to mine a corpus of invariants on which new tools can rely for tasks such as automated patching \cite{perkins2009automatically} and test generation \cite{artzi2006finding, 10.1007/11531142_22}.

There is some related work on judging the validity of invariants, both those Daikon generates and invariants in general. Staats \etal study the accuracy with which programmers can classify automatically generated invariants and find that this accuracy can be remarkably poor; users in their study misclassified between 9\% and 32\% of correct invariants and between 26\% and 59\% of incorrect invariants. We indeed found that classifying real-world invariants is a complicated task, often requiring inspection of the surrounding code. We also relied on two annotators to reduce the risk of mistakes and aimed to set objective guidelines for conflict resolution (see \Cref{sec:golden}), omitting any unresolved invariants entirely.

A few studies have assessed the validity of Daikon's invariants directly. Polikarpova \etal do so specifically on the Eiffel programming language \cite{polikarpova2009comparative}. They also identify invariants in terms of both relevance and validity, using quite similar criteria for both (although we did not assess the validity of irrelevant invariants). They find a strong correlation between the size of the test suite and the validity of generated invariants. However, they also find substantially higher proportions of valid invariants: with a medium-sized test suite, 54-67\% of their pre- and post-conditions are relevant (this is similar to our work), and 83-92\% are valid (this is far higher than we found). This discrepancy is likely due to the programs they consider being much smaller (1.5KLOC at most) than ours as they evaluate only isolated library classes. Kim \& Peterson evaluate Daikon's invariants on larger (C++) systems \cite{kimevaluation}; although their technical report does not include a quantitative evaluation, they do observe many phenomena similar to ours. Among others, they stipulate the high degree of false positives that Daikon produces, even with filters in place, and the absence of semantically insightful invariants, despite the presence of elaborate conditions. Our work further confirms and quantifies their findings and aims to provide a solution in the form of our learning methodology and model.

\section{Summary}
We set out to address a complex question: are real-world method invariants natural? Answering this involves assessing what such invariants look like in practice and how they relate to their methods. We conducted a manual study on a widely used pre- and post-condition inference tool (Daikon) on methods from several large systems. Our findings show that substantial challenges exist for practically useful invariant inference, such as needing better automatic validity assessment, better identification of relevant invariants, and inference of more meaningful complex invariants. In this work, we demonstrate the potential for a learning-based approach that addresses the first challenge. We contribute an automated approach to mining a large corpus of annotated invariants with noisy, yet useful labels, and a model that can learn to meaningfully validate invariants based on this data. Our graph neural network based invariant validator is able to accurately identify valid invariants in our automatically mined corpus, even across projects. Importantly, it generalizes well to our human-annotated corpus, whereas no simpler model came close to this performance. Our results set the stage for learning of properties of a program's behavior from its source code.

\bibliographystyle{ACM-Reference-Format}
\bibliography{references}

\end{document}